# Topological response of the anomalous Hall effect in $MnBi_2Te_4$ due to magnetic canting

S.-K. Bac,[1] K. Koller,[1] F. Lux,[2] J. Wang,[1] L. Riney,[1] K. Borisiak,[1] W. Powers,[1] M. Zhukovskyi,[3] T. Orlova,[3] M. Dobrowolska,[1] J. K. Furdyna,[1] N. R. Dilley,[7] L. P. Rokhinson,[7,8,9] Y. Mokrousov,[2,4] R. J. McQueeney,[5] O. Heinonen,[6] X. Liu,[1] B. A. Assaf[1]

[1] Department of Physics, University of Notre Dame, Notre Dame, IN 46556, USA

[2] Institute of Physics, Johannes Gutenberg University Mainz, 55099 Mainz, Germany

[3] Notre Dame Integrated Imaging Facility, University of Notre Dame, Notre Dame, IN 46556, USA

[4] Peter Grünberg Institut and Institute for Advanced Simulation, Forschungszentrum Jülich and JARA, 52425 Jülich, Germany

[5] Ames Laboratory, Ames, Iowa 50011, USA

[6] Materials Science Division, Argonne National Laboratory, Lemont, Illinois 60439, USA

[7] Birck Nanotechnology Center, Purdue University, West Lafayette, Indiana 47907, USA

[8] Department of Physics and Astronomy, Purdue University, West Lafayette, Indiana 47907, USA

[9] Department of Electrical and Computer Engineering, Purdue University, West Lafayette, Indiana 47907, USA

**Abstract.**

Three-dimensional (3D) compensated $MnBi_2Te_4$ is antiferromagnetic, but undergoes a spin-flop transition at intermediate fields, resulting in a canted phase before saturation. In this work, we experimentally show that the anomalous Hall effect (AHE) in $MnBi_2Te_4$ originates from a topological response that is sensitive to the perpendicular magnetic moment and to its canting angle. Synthesis by molecular beam epitaxy allows us to obtain a large-area quasi-3D 24-layer $MnBi_2Te_4$ with near-perfect compensation that hosts the phase diagram observed in bulk which we utilize to probe the AHE. This AHE is seen to exhibit an antiferromagnetic response at low magnetic fields, and a clear evolution at intermediate fields through surface and bulk spin-flop transitions into saturation. Throughout this evolution, the AHE is super-linear versus magnetization rather than the expected linear relationship. We reveal that this discrepancy is related to the canting angle, consistent with the symmetry of the crystal. Our findings bring to light a topological anomalous Hall response that can be found in non-collinear ferromagnetic, and antiferromagnetic phases.

**Introduction**

Magnetic topological insulators (MTIs) have attracted tremendous attention in the past decade, as they host topological quantum states that emerge when non-trivial band structures are subjected to the Zeeman interactions.[1] These include Weyl fermions,[2,3] the quantum anomalous Hall insulator,[4,5,6] and the axion insulator.[7] MTIs were obtained by doping a topological insulator with Cr or Mn until the recent discovery of intrinsic MTIs that host layers of magnetically ordered transition metals or rare earths.[8,9] $MnBi_2Te_4$ is such an intrinsic MTI. Different magnetic states have been shown to arise in $MnBi_2Te_4$ on demand.[8,10,11,12] This material is a layered two-dimensional antiferromagnet with Mn atoms occupying a separate layer in a septuple layer (SL) structure resembling that of the quintuple in $Bi_2Te_3$ (see Fig. 1(a)). The intralayer magnetic exchange between Mn atoms is ferromagnetic (FM) and dominant. The interlayer exchange is antiferromagnetic (AFM). These yield an AFM ground state with perpendicular anisotropy.

In flake form, in the ultra-thin limit, $MnBi_2Te_4$ has been studied[13,14,7,15] and can either host a compensated AFM state if its thickness amounts to an even number of SLs or an uncompensated FM state if it amounts to an odd number. The former is particularly interesting since, in the AFM state, the bottom surface of this TI will experience a magnetic exchange interaction of opposite sign to that of the top surface, resulting in an axion insulator state.[7] In the bulk limit, $MnBi_2Te_4$ has also been studied and shown to host a surface-spin-flop transition followed by a canted magnetic phase at intermediate magnetic fields.[16,8,17] At high field, when the ferromagnetic state is reached $MnBi_2Te_4$ was argued to host type-II Weyl fermions.[3] However, in the presence of non-collinear and canted magnetic orders[18,19,17,20,21], this material, as well as MTIs in general, can yield exciting undiscovered electronic effects.

In this work, we grow a pure 24-SL $MnBi_2Te_4$ thin film by molecular beam epitaxy (MBE) that hosts a magnetic phase diagram that includes a FM, an AFM and a canted phase, as in bulk. We reveal this phase diagram through the observation of changes in the anomalous Hall effect (AHE). We study the scaling of the AHE in the presence of canting as well as its evolution with temperature through the various magnetic phases hosted by $MnBi_2Te_4$. We show that the canting angle can alter the expected scaling relation of the AHE with magnetization, even in the absence of planar chiral textures such as skyrmions. Beyond previous experimental studies, we experimentally show that an AHE term proportional to the cube of magnetization is needed to account for the observed scaling. We theoretically justify the origin of this term in supplementary section 1. Our results provide an important step in the understanding of non-collinear magnetic orders and how they impact electric transport in MTIs.

**Results**

**Material synthesis**.

$MnBi_2Te_4$ films are synthesized by MBE on GaAs(111)B substrates. The substrates are initially annealed up to 580°C to desorb the native surface oxide. A GaAs buffer layer (50 nm) is then grown to improve the substrate surface quality. This step is critical to obtain a flat interface, and a smooth layer. The GaAs surface is then treated with a Te flux at 580°C to obtain a Te-termination. A $Bi_2Te_3$ buffer layer (4 quintuple layers) is then grown at 280°C. The $Bi_2Te_3$ layer is further annealed at 360°C under a Te flux to further improve surface smoothness. We compare the growth of three samples A, B,and C where we sequentially[22] exposed the substrate to a flux of the following: Mn-Bi-Te (for 30s), Mn-Te (30s for A and C, and 15s for B) and Te (120s for A and B, and 180s for C). This is repeated 20 times, all while maintaining a substrate temperature of 320°C. The growth is carried out under Te rich conditions for all samples.[23] The layers interdiffuse and yield a continuous $MnBi_2Te_4$ layer in sample A, and a $MnBi_2Te_4$-$Bi_2Te_3$ heterostructure in sample B and C likely due to the lower Mn-Te deposition time or longer annealing time.

**Characterization.**

A combination of structural and magnetic characterization allows us to confirm formation of a pure $MnBi_2Te_4$ layer with no evidence of interpenetrating $Bi_2Te_3$. From X-ray diffraction (XRD) measurements shown in Figure 1(b), it is evident sample A hosts strong Bragg peaks characteristic of $MnBi_2Te_4$, while the other samples contain both $Bi_2Te_3$ and $MnBi_2Te_4$. The XRD measurements thus reveal the formation of a pure macroscopic film of $MnBi_2Te_4$ in sample A with a *c*-lattice constant equal to 41.20 Å. Figure 1(c) shows a transmission electron microscope (TEM) image taken on sample A. The image reveals a near-ideal stacking of 24 SLs of $MnBi_2Te_4$ with no evidence of interpenetrating $Bi_2Te_3$ layers seen in previous studies[9,24,25] and in sample B (see supplementary discussion 2). Figure 1(d) compares the remanent magnetization measured using SQUID magnetometry at low magnetic field ($3\ mT$) versus temperature.

A FM transition is observed in the samples containing both $Bi_2Te_3$ and $MnBi_2Te_4$, while sample A only exhibits a slight deviation from the baseline of the measurement. This deviation could result from antisite $Mn_{Bi}$ defects[26]. This comparison strongly indicates the quasi-compensated AFM nature of sample A resulting from the structural homogeneity revealed by TEM and X-ray diffraction measurements.

**Anomalous Hall and magnetic response.**

The Hall effect measured in samples A, B and C at 4.2 K are compared in Fig. 2(a). In samples B and C, a smaller amount of Mn is introduced, and the AHE is dominated by a strong normal n-type Hall response at high magnetic field. The normal Hall effect in sample A is, however, positive and the overall Hall response of this sample is dominated by the AHE. This AHE comes out qualitatively very similar to the magnetization of $MnBi_2Te_4$ single crystals measured in previous works.[27,28] The magnetization of sample A obtained from VSM magnetometry measurements at 2 K is plotted in Fig. 2(b) versus magnetic field along with the anomalous Hall resistance extract by subtracting the high field slope from $R_{xy}$. In both measurements, magnetic regimes can already be identified from discontinuities in the slopes as a function of magnetic field. A total of five regimes are visible from the AHE. Numerical simulations discussed next allow us to understand these regimes.

Magnetic simulations are carried out by utilizing an energy minimization scheme of the modified Mills model and a Monte Carlo approach. In the Mills model,[29,16] the total energy of the system is given by:

$$E = \sum_{i=1}^{N-1} J_i \mathbf{s}_i \cdot \mathbf{s}_{i+1} - \frac{1}{2}\sum_{i=1}^{N} K_i (\mathbf{s}_i \cdot \hat{\mathbf{z}})^2 - \mathbf{H} \cdot \sum_{i=1}^{N} \mathbf{s}_i \quad (1)$$

with the reduction rules, $\mathbf{s}_i = \mathbf{s}_i + \delta_{i=1,N}(\lambda_s - 1)\mathbf{s}_i, J_i = J + \delta_{i=1,N-1}(\lambda_J - 1)J$, and $K_i = K + \delta_{i=1,N}(\lambda_K - 1)K$. $\lambda_A (A = s, J, K)$ represents the reduction of the magnetization $s_i$, of the exchange coupling $J_i$, and of the anisotropy energy $K_i$, respectively. The magnetization was computed from the modified Mills model for $N = 24$ layers with the parameter set $J = 2.35\ T, K = 0.6J, \lambda_s = 0.6, \lambda_J = 0.8$, and $\lambda_K = 0.6$. The model is implemented with open boundary conditions to account for two surfaces and only includes the interaction of nearest-neighboring layers in the vertical direction. In Fig. 2(c), we plot the perpendicular component of magnetization obtained from the Mills model versus magnetic field (red curve). Classical Monte Carlo simulations[30,31,32] at 2 K, as a function of increasing field, are also performed for a homogeneous 24-layer film with free surfaces. The exchange and anisotropy parameters are shown in the methods section. The single-ion anisotropy parameter is slightly reduced compared to ref. [30] to better align the spin-flop fields with experimental data and Mills model. The Monte Carlo simulations include an intralayer exchange interaction unlike the Mills model. The implementation of both models is discussed in the methods section. The simulations reproduce the behavior of the modified Mills model above $3\ T$. They are shown as the blue curve in Fig. 2(c).

The spin texture obtained from Monte Carlo simulations allows us to identify a variety of magnetic phases arising as the field is swept. In Fig. 2(d), we plot the spin texture of the 24-layer system versus magnetic field between $0$ and $10\ T$. At low fields, in region (i), the system is a collinear antiferromagnet (AFM) and yields the AFM Hall plateau observed in Fig. 2(b). Up to 3.4T, in region (ii), the surface layer that is antiparallel with respect to field starts to flip yielding a positive slope in the magnetization versus field plot shown in Fig. 2(c) and corresponds to a change in the observed AHE slope in Fig. 2(b). In the modified Mills model, this transition is abrupt and yields a sudden surface spin flop state (SSF). The Monte Carlo simulation at 2K results in a smeared transition due to thermal fluctuations, yielding a progressive spin flip with increasing field. At $3.4\ T$, in region (iii), the bulk spin flop transition (BSF) occurs yielding a sudden jump in the AHE observed in fig. 2(b). While region (iii) appears as an abrupt BSF transition in the simulation (Fig. 2(d)), it is broadened in the experiment as various domains in the sample go through this transition at slightly different fields. Afterwards in region (iv) the magnetization is canted and slowly rotates towards the z-axis. This is the canted antiferromagnetic (CAFM) phase identified in previous works.[33,17] In Fig. 2(b), a linear increase of the AHE is observed in region (iv). Above $8\,T$, the system is a saturated ferromagnet (region (v)). While $R_{xy}$ is far from being quantized (<< h/e$^2$) in this region, the Hall conductance $G_{xy}$ extracted from the Drude tensor saturates close to 0.2e$^2$/h above $8\,T$.

Despite the remarkable agreement of the magnetic simulations with the AHE, the magnetization exhibits some differences at low magnetic field. The remanent magnetic hysteresis loop at low field seen in Fig. 2(b,c) (and in supplementary discussion 3 for the AHE) may be due to a defect state arising from antisites[26,34,35] in the structure or a disordered surface layer. The Monte Carlo simulations do yield a remanent surface magnetization, however, the measured relative remanence (M(0T)/M(7T)) is larger than what is seen in calculations. Between $3.4\,T$ and 7T, the measurements and simulations converge at most within two standard deviations , but the magnetization saturates close to $4\mu_B/u.c.$ lower than the maximum expected for $MnBi_2Te_4$. This could be due to $Mn_{Bi}$ antisites that were shown to couple antiferromagnetically to the Mn layers, yielding a drop in the net moment. However, these Mn atoms are not expected to yield a field dependent magnetization at the fields of interest.[35]

**Temperature dependence and AHE phase diagram.**

In Fig. 3(a), we plot the temperature dependence of the Hall resistance from sample A. The magnetoresistance is shown in supplementary discussion 4. The Hall resistance allow us to construct a magnetic phase diagram, from electrical measurements. We take the first derivative of the $R_{xy}$ versus $B$ data shown Fig. 3(a) and plot it in Fig. 3(b) to evidence more strongly the temperature dependence of each magnetic transition. The resulting magnetic phase diagram is shown in Fig. 3(c) and agrees with previous work on single crystals.[17] As can be seen in that figure, the onset of each magnetic regime decreases with increasing temperature. Particularly, we can see a suppression of the AFM phase close to 15 K. Between 15 and 20 K, the canted surface and bulk magnetic phases remain present. Above 20 K, the Hall response remains non-linear, but the discrete slope changes observed at low temperature disappear, indicating that the material enters a paramagnetic/ferromagnetic phase.

The temperature dependent measurements also elucidate a possible ambipolar behavior at high magnetic field as the slope of the Hall effect is seen to change with increasing temperature. We hypothesize that sample A hosts coexisting electrons and holes possibly from Mn acting as an acceptor if it substitutes for Bi.

**Scaling of the AHE.**

We next investigate the scaling of the AHE with magnetization to understand the impact of the magnetic structure of $MnBi_2Te_4$ on its AHE. We focus on the canted AFM regime, for which the magnetization and magnetic simulations show a good agreement. Generally speaking, the intrinsic AHE in magnetic materials is given by [36]

$$\rho_{xy}^{A} = a\rho_{xx}^{2}M \quad (2)$$

Here, $a$ is a coefficient proportional to the Berry curvature, $\rho_{xy}^{A}$ is the anomalous Hall resistivity and $\rho_{xx}$ is the longitudinal resistivity. $M$ is the magnetization. Thus, the Hall conductance $\sigma_{xy}^{A} = \rho_{xy}^{A}/\rho_{xx}^{2}$ is simply proportional to $M$. Even in the presence of Weyl nodes expected for this material, the Hall conductance is proportional to the node separation,[2] which in turn is proportional to the magnetization. If the field dependence of $\rho_{xx}$ is small such as in our case, then both $R_{xy}^{A}$ and $\sigma_{xy}^{A}$ should be linear versus $M$. In Fig.4(a), we plot $\sigma_{xy}^{A}$ versus $M$ for different temperatures. We restrict this analysis to the shaded region in Fig. 2(c), for which the magnetization and the simulations agree the most (within error). The uncertainty on the magnetization is also included in the scaling analysis shown in Fig. 4(a) (see supplementary discussion 5). While at 20 to 30 K the $\sigma_{xy}^{A}$ is seen to scale almost linearly with $M$, we recover a remarkable change in this scaling relation at low temperature. At temperatures where the material hosts a canted phase, the AHE starts to exhibit a super-linear scaling with magnetization.

This super-linear behavior violates the established scaling relation of the AHE. The "excess" Hall effect resembles what is observed in non-collinear magnetic systems such $SrRuO_3$ [37,38] and in materials that host magnetic skyrmions.[39] The crystal structure of $MnBi_2Te_4$ is centrosymmetric, thus reducing its likelihood to host skyrmions.[40] Note that a recent work has investigated a similar "excess" AHE in MBE grown $MnTe_2/MnBi_2Te_4$,[41] but did not reach sufficiently high fields to measure the AHE in the canted phases.

To account for the non-linear scaling, we model the anomalous Hall conductance by adding an additional unexpected component due to canting. The Berry curvature driven response of the AHE due to canted magnetism has only been recently investigated.[42,43,44] Theoretical studies suggest that additional AHE components - referred to as a chiral and a crystal AHE - can arise depending on crystallographic symmetry of the material (supplementary discussion 1).[42,45,43] For $MnBi_2Te_4$, we determine the symmetry-allowed contributions to the AHE up to third order in the out-of-plane ferromagnetic moment $M$ as well as in the in-plane antiferromagnetic moment $\mathbf{M}_- = \mathbf{M}_{\boldsymbol{A}} - \mathbf{M}_{\boldsymbol{B}}$. Here $\mathbf{M}_{\boldsymbol{A},\boldsymbol{B}}$ are the magnetic moment of two Mn sublattices. We find[46,47]

$$\sigma_{xy}^{A} = \gamma_{AHE}\frac{M}{M_{sat}} + \gamma_{nAHE}\frac{M^3}{M_{sat}^3} + \gamma_{\chi}\frac{MM_-^2}{M_{sat}^3}, \quad (3)$$

when assuming the space group symmetry R-3m.[40,48] The series coefficients $\gamma_i$ are material parameters. Under the constraint that the magnitude of the individual magnetic moments is fixed, one can write $M_-^2 = M_{sat}^2 - M^2$. This expansion is therefore able to justify why $\sigma_{xy}^{A} \sim M^{\alpha}$ with $1 \leq \alpha \leq 3$ is a good approximation. Further, Fig. 4 (a) demonstrates that approximately $\gamma_{nAHE} = 0$, since $\alpha = 1$ is close to the exact exponent in the ferromagnetic phase for $T > 20$K. The anomalous scaling behavior in the CAFM phase could then be related to a finite $\gamma_\chi$. The term $MM_-^2$ can be reformulated in a way, which makes its physical content more apparent. Namely, we introduce the vector chirality $\boldsymbol{\chi} = \mathbf{M}_- \times \mathbf{M}$ and consider the experimentally relevant case, where $\mathbf{M} = M\mathbf{e}_z$ and $\mathbf{M}_- = M_-\mathbf{e}_x$. Then, one can write the additional Hall contribution in terms of the chirality in two ways as

$$MM_-^2 = -\chi_y\, M_- = \frac{\chi_y^2}{M} \quad (4).$$

Near the purely antiferromagnetic state this anomalous Hall contribution is therefore linear in the components of the vector chirality and the canting is introduced via the small ferromagnetic component. By definition, it therefore classifies as a so-called *chiral Hall effect*.[42] In the vicinity of the ferromagnetic state, the canting enters via a small antiferromagnetic component and the effect is seen to be quadratic in the chirality (supplementary discussion 1). This can be seen as the defining quality of a *crystal Hall effect* for canted ferromagnets.[42]

Overall, $M\,M_-^2$ is a manifestly canting-driven contribution to the AHE, *not* proportional to the overall magnetization. To compare with the experiment, we write

$$\frac{\sigma_{xy}^A(m)}{\sigma_{xy}^A(1)} \sim m + \frac{\gamma_\chi}{\gamma_{AHE}} m\,(1-m^2) \quad (5)$$

The non-linear scaling exponent is therefore controlled by the relative magnitude of the canting contributions $m = M/M_{sat} = \cos(\phi)$ where $\phi$ is the canting angle with respect to the vertical. Indeed, canting is clearly visible in the spin texture shown in Fig. 2(d) between $3.8\,T$ and $7.8\,T$. In Fig. 4(b), we compare our calculated anomalous Hall conductivity (Eq. 3) to the experimental data. The red curve uses $\gamma_{AHE} = 2.55\ (\Omega.\mathrm{T.cm})^{-1}$, $\gamma_\chi = -0.75\ (\Omega.\mathrm{T.cm})^{-1}$ and the blue curve uses $\gamma_\chi = 0$. It is evident that a nonlinear contribution is required to account for the difference in slope between the magnetization obtained from the Mills model and the AHE. To get a better comparison with the magnetization data, we also plot the scaling relation obtained by comparing the calculated anomalous Hall resistance to the simulated magnetization. A super-linear scaling relation with $M^{1.3}$ is reproduced in the canted phase (Fig. 4(c)). We have hence shown that the AHE in $MnBi_2Te_4$ can be a direct function of the canting angle.

**Discussion**

We have studied the AHE in a 24-SL $MnBi_2Te_4$ layer obtained by MBE and shown that the anomalous Hall response of topological origin in this material contains a canting angle contribution at low temperature. A symmetry analysis has allowed us to observe an AHE scaling that comprises a remarkable contribution that scales cubically with magnetization. This explains the origin of the non-linear scaling of the Hall conductivity with magnetization observed in the experiment. We show that its origin is related to the canting of the magnetic moment induced by the bulk spin flop transition before saturation. Specifically, we have revealed that the AHE measured here can be explained by the first non-trivial, canting-dependent correction which is allowed by the crystallographic symmetry of $MnBi_2Te_4$. A detailed symmetry analysis showing the origin of this effect can be found in supplementary discussion 1. In referring to prior theoretical work, this term has been identified as the chiral Hall effect of canted antiferromagnets. [42,43] By elucidating these effects in $MnBi_2Te_4$, we have explained the origin of the unusual scaling of the Hall effect when canting is present.

**Methods.**

**X-ray diffraction**. X-ray diffraction measurements are carried out at room temperature in a Bruker D8 Discover diffractometer equipped with Cu-Kα-source.

**Magnetometry.** SQUID magnetometry is carried out in a Quantum Design MPMS, down to 4.2 K at various magnetic fields up to $7\,T$. The field is applied perpendicular to the sample plane. The diamagnetism of the GaAs substrate assumed to contribute a linear slope between 6T and 7T is subtracted at each temperature. VSM magnetometry is carried out in a Quantum Design PPMS system equipped with a VSM head up to 9T. The diamagnetism of the GaAs substrate is subtracted at each temperature.

**Transmission electron microscopy.** High-resolution cross-sectional TEM images were acquired using a double tilt holder and Titan 80-300 transmission electron microscope (Thermo Fisher Scientific, USA) equipped with a field emission gun, operated at 300 kV. STEM images were acquired using a high-angle, annular dark field detector (HAADF) and bright field detector (Fischione Instruments). For compositional analysis, energy-dispersive X-ray spectroscopy (EDS) maps were obtained in STEM mode using the Ultim Max TLE EDS system (Oxford Instruments) equipped with a large solid angle silicon drift detector. TEM samples were prepared by focused ion beam etching using the standard lift-out technique.

**Electrical transport**. Electrical Hall effect and magnetoresistance measurements are carried out in an Oxford Instruments cryostat up to 16 $T$ and down to 1.4 K. The excitation current is maintained at 100 $\mu A$. Rectangular samples cleaved from the GaAs wafer are measured in a 5-wire Hall configuration. The Hall conductivity is extracted as follows:

$$\sigma_{xy} = \frac{\rho_{xy}}{\rho_{xx}^2 + \rho_{xy}^2} \quad (6)$$

$$\rho_{xy} = R_{xy} t \;\; and \;\; \rho_{xx} = \frac{R_{xx} w}{L} t \;\; (7)$$

$w, L \; and \; t$ are the sample width, length, and thickness respectively. For sample A, the measurements are carried out on a rectangular piece with $w \times L = 1.05 \, mm^2$. $\sigma_{xy}^A$ utilizes the anomalous Hall resistance signal after a linear Hall background is removed at high magnetic field from $R_{xy}$.

**Mills model.** The red magnetization curve in Fig. 2(c) was obtained by using the revised Mills model shown in Eq. (1) with N = 24. The ground state at a positive high field is searched by comparing total energies of spin configurations relaxed from 100 initial random configurations. After that, each sampling points are searched from previous configurations. In this model, we did not consider the effect of thermal fluctuation and the ferromagnetic intralayer exchange coupling.

**Monte Carlo model.** The blue magnetization curve in Fig. 2(c) was obtained by using a Monte Carlo simulation[31,32] similar to the ones carried out in ref [30]. The simulation includes an intralayer ferromagnetic exchange coupling between Mn nearest-neighbors within the same layer. The calculations are carried out at 2K and include the effect of thermal fluctuations. The following parameters are used during the simulation as they yielded the best agreement with the experimental data: $m = 5\mu_B/Mn, , J_{inter} = -0.0081mRy\ , J_{intra} = 0.03mRy, K = -0.01mRy,$ system size: (11x11 Mn in layer) x (24 layers).

### Data availability statement

The data that support the findings of this study are available from the corresponding author upon reasonable request.

**Acknowledgements.** We acknowledge support from the National Science Foundation grant NSF-DMR-1905277. R.J.M. and O.H. were supported by the Center for Advancement of Topological Semimetals, an Energy Frontier Research Center funded by the US Department of Energy Office of Science, Office of Basic Energy Sciences, through the Ames Laboratory under Contract No. DE-AC02-07CH11358. Y.M. and F.L. acknowledge support from Deutsche Forschungsgemeinschaft (DFG, German Research Foundation) - TRR 173 - 268565370 (project A11), TRR 288 – 422213477 (project A06), and project MO 1731/10-1. LPR acknowledges supported by the U.S. Department of Energy, Office of Science, National Quantum Information Science Research Centers, Quantum Science Center.

**Author contributions.** SKB performed magnetotransport and X-ray diffraction measurements. XL synthesized the samples with assistance from JW. KK and SKB performed SQUID magnetometry. KB and SKB performed numerical calculations using the Mills model. LR, JW and WP assisted in X-ray diffraction,

SQUID measurements and transport measurements. RJM and OH carried out Monte Carlo simulations. FL and YM carried the symmetry analysis and derived additional contributions to the Hall effect. MZ and TO prepared and carried out TEM measurements. ND and LPR carried out VSM measurements. SKB, XL and BAA conceived the experiments and jointly analyzed the results. MD, JKF, XL and BAA supervised the project and provided input on the analysis. All authors contributed to the writing of the manuscript.

**Competing Interest**

The authors declare that there are no non-financial and financial competing interests.

**References**

1. Tokura, Y., Yasuda, K. & Tsukazaki, A. Magnetic topological insulators. *Nat. Rev. Phys.* **1**, 126–143 (2019). http://www.nature.com/articles/s42254-018-0011-5

2. Burkov, A. A. & Balents, L. Weyl semimetal in a topological insulator multilayer. *Phys. Rev. Lett.* **107**, 127205 (2011). http://link.aps.org/doi/10.1103/PhysRevLett.107.127205

3. Lee, S. H. *et al.* Evidence for a magnetic-field-induced ideal type-II Weyl state in antiferromagnetic topological insulator Mn(Bi1-xSbx)2Te4. *Phys. Rev. X* **11**, 031032 (2021). https://link.aps.org/doi/10.1103/PhysRevX.11.031032

4. Chang, C.-Z. *et al.* Experimental observation of the quantum anomalous Hall effect in a magnetic topological insulator. *Science (80-. ).* **340**, 167–70 (2013). https://www.science.org/doi/10.1126/science.1234414

5. Chang, C.-Z. *et al.* High-precision realization of robust quantum anomalous Hall state in a hard ferromagnetic topological insulator. *Nat. Mater.* **14**, 473–477 (2015). http://www.nature.com/articles/nmat4204

6. Checkelsky, J. G. *et al.* Trajectory of the anomalous Hall effect towards the quantized state in a

ferromagnetic topological insulator. *Nat Phys* **10**, 731–736 (2014). http://www.nature.com/articles/nphys3053

7. Liu, C. *et al.* Robust axion insulator and Chern insulator phases in a two-dimensional antiferromagnetic topological insulator. *Nat. Mater.* **19**, 522–527 (2020). http://www.nature.com/articles/s41563-019-0573-3

8. Otrokov, M. M. *et al.* Prediction and observation of an antiferromagnetic topological insulator. *Nature* **576**, 416–422 (2019). http://www.nature.com/articles/s41586-019-1840-9

9. Rienks, E. D. L. *et al.* Large magnetic gap at the Dirac point in Bi2Te3/MnBi2Te4 heterostructures. *Nature* **576**, 423–428 (2019). http://www.nature.com/articles/s41586-019-1826-7

10. Li, J. *et al.* Intrinsic magnetic topological insulators in van der Waals layered MnBi 2 Te 4 -family materials. *Sci. Adv.* **5**, eaaw5685 (2019). https://advances.sciencemag.org/lookup/doi/10.1126/sciadv.aaw5685

11. Otrokov, M. M. *et al.* Unique thickness-dependent properties of the van der Waals interlayer antiferromagnet MnBi2Te4 films. *Phys. Rev. Lett.* **122**, 107202 (2019). https://doi.org/10.1103/PhysRevLett.122.107202

12. Zhang, D. *et al.* Topological Axion states in the magnetic insulator MnBi2Te4 with the quantized magnetoelectric effect. *Phys. Rev. Lett.* **122**, 206401 (2019). https://doi.org/10.1103/PhysRevLett.122.206401

13. Ovchinnikov, D. *et al.* Intertwined topological and magnetic orders in atomically thin Chern insulator MnBi2Te4. *Nano Lett.* **21**, 2544–2550 (2021). https://pubs.acs.org/doi/10.1021/acs.nanolett.0c05117

14. Deng, Y. *et al.* Quantum anomalous Hall effect in intrinsic magnetic topological insulator MnBi 2

Te 4. *Science (80-. ).* **367**, 895–900 (2020). https://www.sciencemag.org/lookup/doi/10.1126/science.aax8156

15. Gao, A. *et al.* Layer Hall effect in a 2D topological axion antiferromagnet. *Nature* **595**, 521–525 (2021). http://www.nature.com/articles/s41586-021-03679-w

16. Sass, P. M., Kim, J., Vanderbilt, D., Yan, J. & Wu, W. Robust A-type order and spin-flop transition on the surface of the antiferromagnetic topological insulator MnBi2Te4. *Phys. Rev. Lett.* **125**, 037201 (2020). https://link.aps.org/doi/10.1103/PhysRevLett.125.037201

17. Lee, S. H. *et al.* Spin scattering and noncollinear spin structure-induced intrinsic anomalous Hall effect in antiferromagnetic topological insulator MnBi2Te4. *Phys. Rev. Res.* **1**, 012011 (2019). https://link.aps.org/doi/10.1103/PhysRevResearch.1.012011

18. Paul, N. & Fu, L. Topological magnetic textures in magnetic topological insulators. *Phys. Rev. Res.* **3**, 033173 (2021). https://link.aps.org/doi/10.1103/PhysRevResearch.3.033173

19. Xiao, C., Tang, J., Zhao, P., Tong, Q. & Yao, W. Chiral channel network from magnetization textures in two-dimensional MnBi2Te4. *Phys. Rev. B* **102**, 125409 (2020). https://link.aps.org/doi/10.1103/PhysRevB.102.125409

20. Zhang, R. X., Wu, F. & Das Sarma, S. Möbius insulator and higher-order topology in MnBi2nTe3n+1. *Phys. Rev. Lett.* **124**, 136407 (2020). https://doi.org/10.1103/PhysRevLett.124.136407

21. Puphal, P. *et al.* Topological magnetic phase in the candidate Weyl semimetal CeAlGe. *Phys. Rev. Lett.* **124**, 17202 (2020). https://doi.org/10.1103/PhysRevLett.124.017202

22. Gong, Y. *et al.* Experimental realization of an intrinsic magnetic topological insulator. *Chinese Phys. Lett.* **36**, 076801 (2019). https://iopscience.iop.org/article/10.1088/0256-

307X/36/7/076801

23. Zhu, K. *et al.* Investigating and manipulating the molecular beam epitaxy growth kinetics of intrinsic magnetic topological insulator MnBi2Te4with in situ angle-resolved photoemission spectroscopy. *J. Phys. Condens. Matter* **32**, (2020). https://iopscience.iop.org/article/10.1088/1361-648X/aba06d

24. Hagmann, J. A. *et al.* Molecular beam epitaxy growth and structure of self-assembled Bi2Se3/Bi2MnSe4 multilayer heterostructures. *New J. Phys.* **19**, 85002 (2017). https://doi.org/10.1088/1367-2630/aa759c

25. Deng, H. *et al.* High-temperature quantum anomalous Hall regime in a MnBi2Te4/Bi2Te3 superlattice. *Nat. Phys.* **17**, 36–42 (2021). http://www.nature.com/articles/s41567-020-0998-2

26. Lee, J. S. *et al.* Ferromagnetism and spin-dependent transport in n -type Mn-doped bismuth telluride thin films. *Phys. Rev. B* **89**, 174425 (2014). https://link.aps.org/doi/10.1103/PhysRevB.89.174425

27. Yan, J.-Q. *et al.* Crystal growth and magnetic structure of MnBi2Te4. *Phys. Rev. Mater.* **3**, 064202 (2019). https://link.aps.org/doi/10.1103/PhysRevMaterials.3.064202

28. Chen, B. *et al.* Intrinsic magnetic topological insulator phases in the Sb doped MnBi2Te4 bulks and thin flakes. *Nat. Commun.* **10**, 4469 (2019). http://www.nature.com/articles/s41467-019-12485-y

29. Mills, D. L. Surface spin-flop state in a simple antiferromagnet. *Phys. Rev. Lett.* **20**, 18–21 (1968). https://link.aps.org/doi/10.1103/PhysRevLett.20.18

30. Lei, C., Heinonen, O., MacDonald, A. H. & McQueeney, R. J. Metamagnetism of few-layer topological antiferromagnets. *Phys. Rev. Mater.* **5**, 064201 (2021).

https://link.aps.org/doi/10.1103/PhysRevMaterials.5.064201

31. Skubic, B., Hellsvik, J., Nordström, L. & Eriksson, O. A method for atomistic spin dynamics simulations: implementation and examples. *J. Phys. Condens. Matter* **20**, 315203 (2008). https://iopscience.iop.org/article/10.1088/0953-8984/20/31/315203

32. Evans, R. F. L. *et al.* Atomistic spin model simulations of magnetic nanomaterials. *J. Phys. Condens. Matter* **26**, 103202 (2014). https://iopscience.iop.org/article/10.1088/0953-8984/26/10/103202

33. Sass, P. M. *et al.* Magnetic Imaging of Domain Walls in the Antiferromagnetic Topological Insulator MnBi 2 Te 4. *Nano Lett.* **20**, 2609–2614 (2020). https://pubs.acs.org/doi/10.1021/acs.nanolett.0c00114

34. Wimmer, S. *et al.* Mn-Rich MnSb 2 Te 4 : A Topological Insulator with Magnetic Gap Closing at High Curie Temperatures of 45–50 K. *Adv. Mater.* **33**, 2102935 (2021). https://doi.org/10.1103/PhysRevB.104.064401

35. Lai, Y., Ke, L., Yan, J., McDonald, R. D. & McQueeney, R. J. Defect-driven ferrimagnetism and hidden magnetization in MnBi2Te4. *Phys. Rev. B* **103**, 184429 (2021). https://link.aps.org/doi/10.1103/PhysRevB.103.184429

36. Nagaosa, N., Sinova, J., Onoda, S., MacDonald, A. H. & Ong, N. P. Anomalous Hall effect. *Rev. Mod. Phys.* **82**, 1539–1592 (2010). https://link.aps.org/doi/10.1103/RevModPhys.82.1539

37. Kimbell, G. *et al.* Two-channel anomalous Hall effect in SrRuO3. *Phys. Rev. Mater.* **4**, 054414 (2020). https://link.aps.org/doi/10.1103/PhysRevMaterials.4.054414

38. Mathieu, R. *et al.* Scaling of the anomalous Hall effect in Sr1-xCaxRuO3. *Phys. Rev. Lett.* **93**, 016602 (2004). https://link.aps.org/doi/10.1103/PhysRevLett.93.016602

39. Neubauer, A. *et al.* Topological Hall effect in the A-phase of MnSi. *Phys. Rev. Lett.* **102**, 186602 (2009). https://link.aps.org/doi/10.1103/PhysRevLett.102.186602

40. Zeugner, A. *et al.* Chemical aspects of the candidate antiferromagnetic topological insulator MnBi2Te4. *Chem. Mater.* **31**, 2795–2806 (2019). https://pubs.acs.org/doi/10.1021/acs.chemmater.8b05017

41. Tai, L. *et al.* Polarity-tunable anomalous Hall effect in magnetic topological insulator MnBi2Te4. *Preprint at https://arxiv.org/abs/*2103.09878 (2021).

42. Kipp, J. *et al.* The chiral Hall effect in canted ferromagnets and antiferromagnets. *Commun. Phys.* **4**, 99 (2021). http://www.nature.com/articles/s42005-021-00587-3

43. Lux, F. R., Freimuth, F., Blügel, S. & Mokrousov, Y. Chiral Hall effect in noncollinear magnets from a cyclic cohomology approach. *Phys. Rev. Lett.* **124**, 096602 (2020). https://link.aps.org/doi/10.1103/PhysRevLett.124.096602

44. Chen, H., Niu, Q. & MacDonald, A. H. Anomalous Hall effect arising from noncollinear antiferromagnetism. *Phys. Rev. Lett.* **112**, 017205 (2014). https://link.aps.org/doi/10.1103/PhysRevLett.112.017205

45. Šmejkal, L. *et al.* Crystal time-reversal symmetry breaking and spontaneous Hall effect in collinear antiferromagnets. *Sci. Adv.* **6**, (2020). https://www.science.org/doi/10.1126/sciadv.aaz8809

46. Birss, R. R. *Symmetry and Magnetism Vol. 3*. (Elsevier Science and Technology, 1966).

47. Dresselhaus, M. S., Dresselhaus, G. & Jorio, A. *Group theory*. (Springer Science & Business Media, 2008). doi:10.1007/978-3-540-32899-5.

48. Yan, J.-Q. *et al.* Crystal growth and magnetic structure of MnBi2Te4. *Phys. Rev. Mater.* **3**, 064202

(2019). https://link.aps.org/doi/10.1103/PhysRevMaterials.3.064202

**Figure Captions**

**Figure 1**. **Characterization of $MnBi_2Te_4$ thin films** (a) Crystal structure of $MnBi_2Te_4$. Magnetic moments of individual Mn atoms are induced by the black arrows. SL: septuple layer (b) X-ray diffraction patterns taken on samples A (red), B (blue), and C (green). The expected Bragg peaks for $MnBi_2Te_4$ and $Bi_2Te_3$ are shown below the data. HS stands for heterostructure (of $Bi_2Te_3$ and $MnBi_2T_4$). (c) Bright field TEM image of the 24-SL sample A. Red lines are a guide for the eye highlighting the layer stacking. (d) Magnetization of samples A (red), B (blue) and C (green) versus temperature measured at a magnetic field of B=3 mT applied along the c-axis.

**Figure 2. Anomalous Hall resistance, magnetization and magnetic structure of 24-SL $MnBi_2Te_4$.** (a) $R_{xy}$ measured in samples A, B and C at 4.2 K. (b). Anomalous Hall resistance $R_{xy}^A$ (red at 1.4 K) measured in sample A compared to its magnetization (blue at 2 K). Labels (i-v) mark 5 different magnetic regimes discussed in the text. $R_{xy}^A$ is obtained by subtracting the high field linear Hall effect from $R_{xy}$. (c) Magnetization measured in sample A at 2 K and 5K using two different instruments (see methods) compared with calculated magnetization using the modified Mills model (red) and Monte Carlo simulations (blue). (d) Evolution of the spin texture of the 24-SL system as a function of magnetic field, obtained from Monte Carlo simulations at 2K. Regions (i)-(v) correspond to the 5 magnetic regimes

observed in (b). The color of the arrows represents the magnitude of the z-component (fully up: red, fully down: blue).

**Figure 3. AHE Phase diagram**. (a) Hall resistance $R_{xy}$ versus magnetic field measured at different temperatures for sample A. (b) First derivative of $R_{xy}$ from sample A with respect to magnetic field at different temperatures. (c) Phase diagram resulting from the transitions observed in the AHE. AFM: antiferromagnetism, SSF: antiparallel surface spin flop transition, BSF: Bulk spin-flop transition. CAFM: canted antiferromagnetism, FM: ferromagnetism.

**Figure 4. Scaling of the AHE in $MnBi_2Te_4$.** (a) Scaling relation between the anomalous Hall conductivity and the magnetization at various temperatures. Data from SQUID magnetometry measured between 2.8T and 7T is used. The shaded red area accounts for a scaling envelope that spans the width of the error bars. The magnetization curves used here are shown in supplementary discussion 5. (b) Anomalous Hall conductivity data (black) compared to Monte Carlo simulation with (red) and without (blue) the canting contribution. (c) Scaling relation between the anomalous Hall conductivity and the magnetization obtained from the Monte Carlo simulation with (red) and without (blue) the canting.

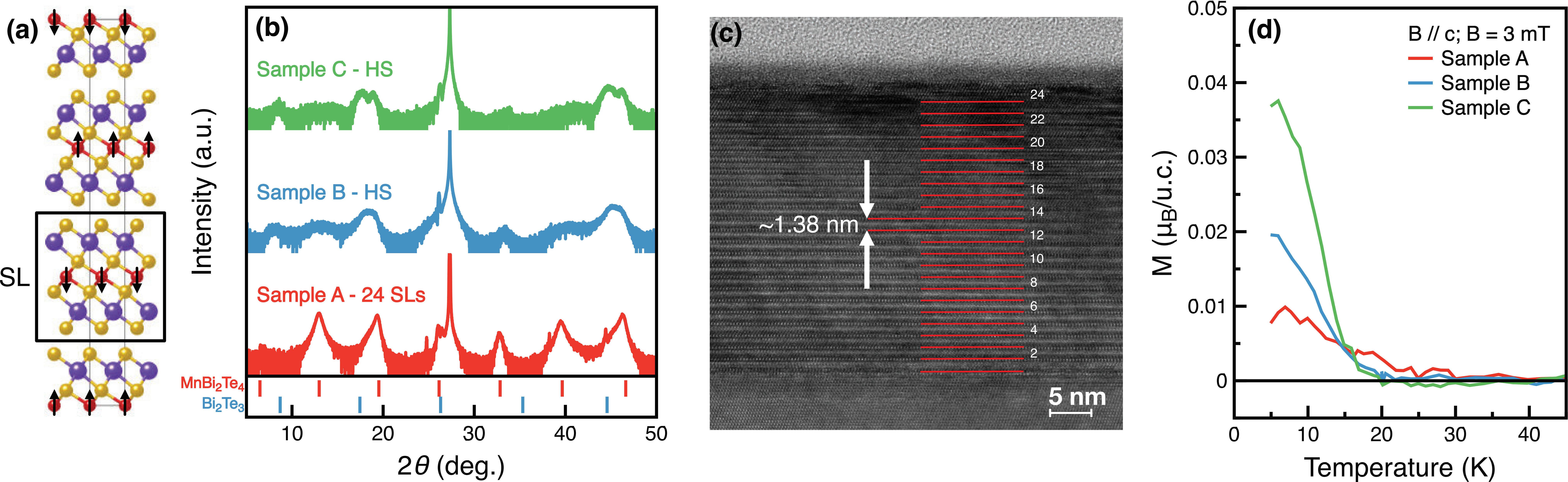

(a)
SL
(b)
Sample C - HS
Sample B - HS
Sample A - 24 SLs
Intensity (a.u.)
MnBi2Te4
Bi2Te3
10
20
30
40
50
2θ (deg.)
(c)
~1.38 nm
24
22
20
18
16
14
12
10
8
6
4
2
5 nm
(d)
B // c; B = 3 mT
Sample A
Sample B
Sample C
M (μB/u.c.)
0.05
0.04
0.03
0.02
0.01
0
Temperature (K)
0
10
20
30
40

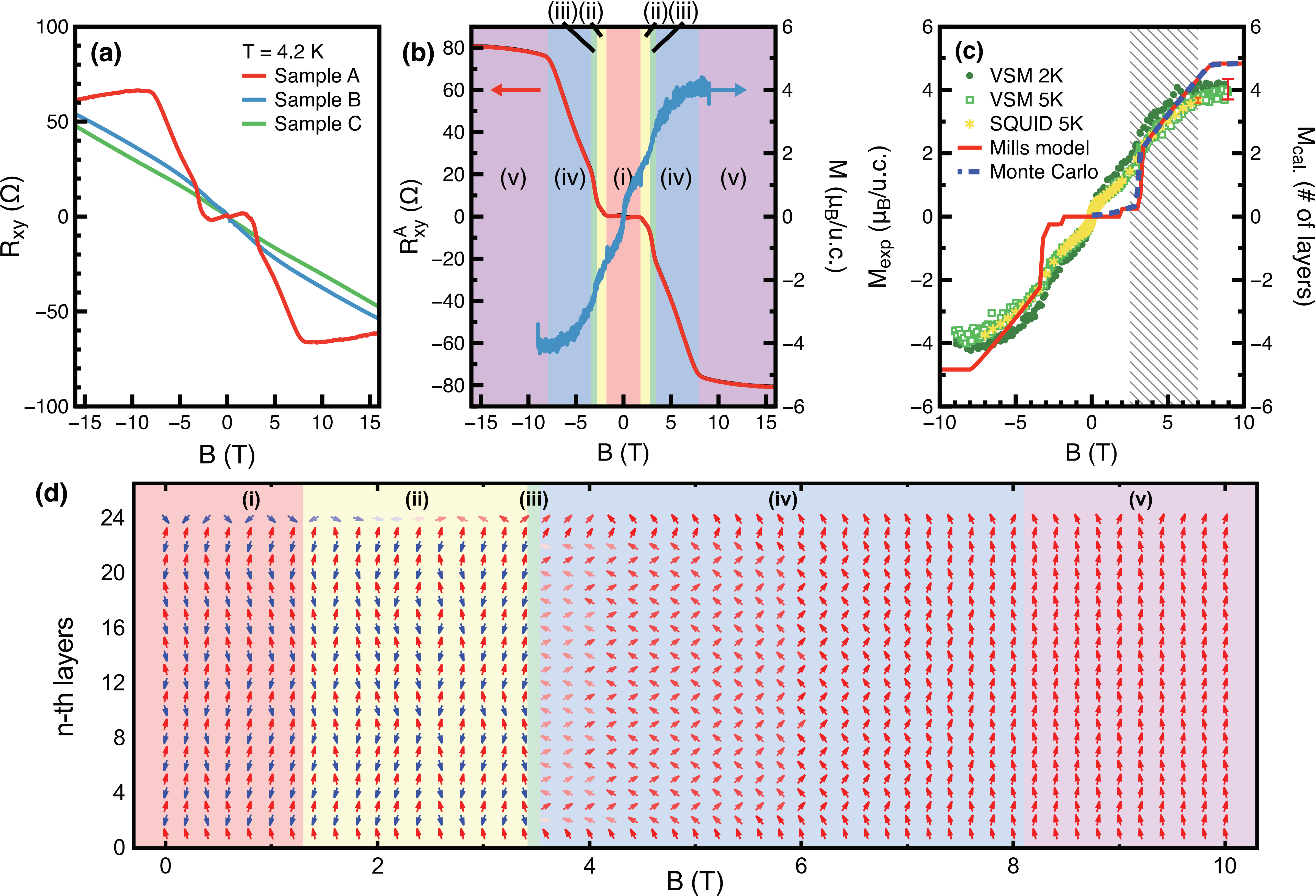

(a)
T = 4.2 K
Sample A
Sample B
Sample C
R_xy (Ω)
B (T)
(b)
(iii)(ii)
(ii)(iii)
(v)
(iv)
(i)
(iv)
(v)
R^A_xy (Ω)
M (μB/u.c.)
B (T)
(c)
VSM 2K
VSM 5K
SQUID 5K
Mills model
Monte Carlo
M_exp (μB/u.c.)
M_cal. (# of layers)
B (T)
(d)
(i)
(ii)
(iii)
(iv)
(v)
n-th layers
B (T)

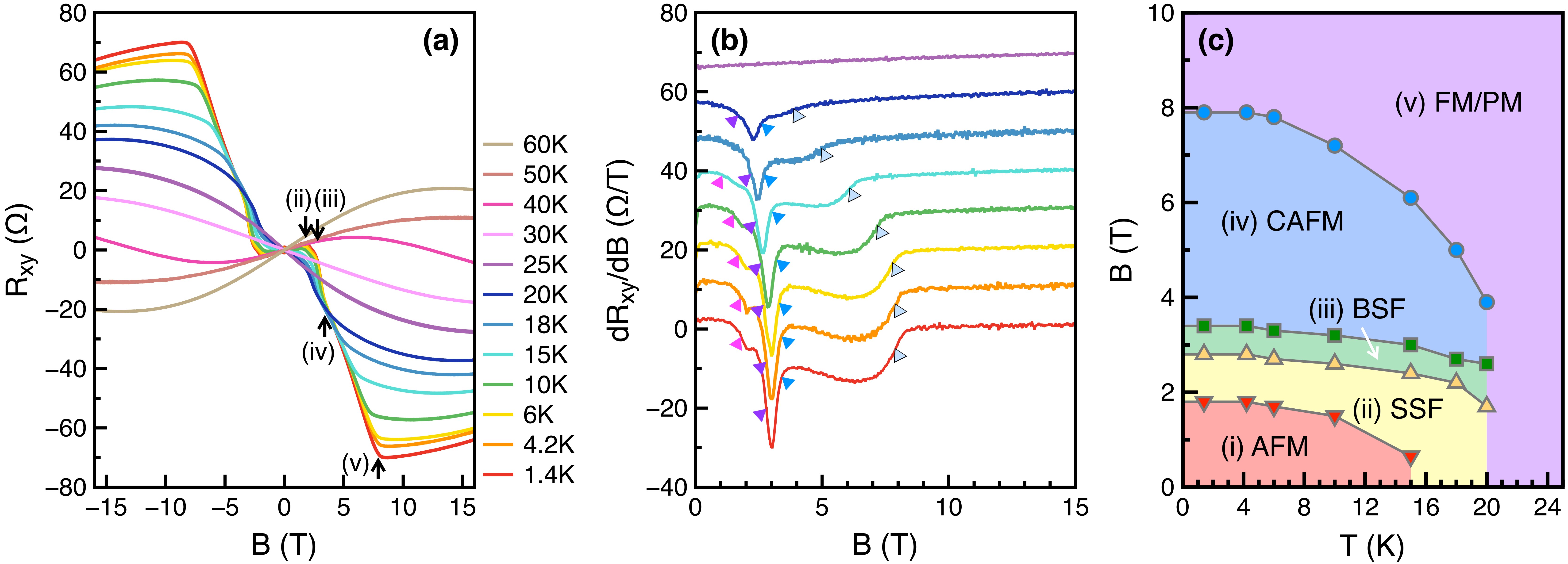
(a)
(ii)
(iii)
(iv)
(v)
60K
50K
40K
30K
25K
20K
18K
15K
10K
6K
4.2K
1.4K
R_xy (Ω)
B (T)
(b)
dR_xy/dB (Ω/T)
B (T)
(c)
(v) FM/PM
(iv) CAFM
(iii) BSF
(ii) SSF
(i) AFM
B (T)
T (K)

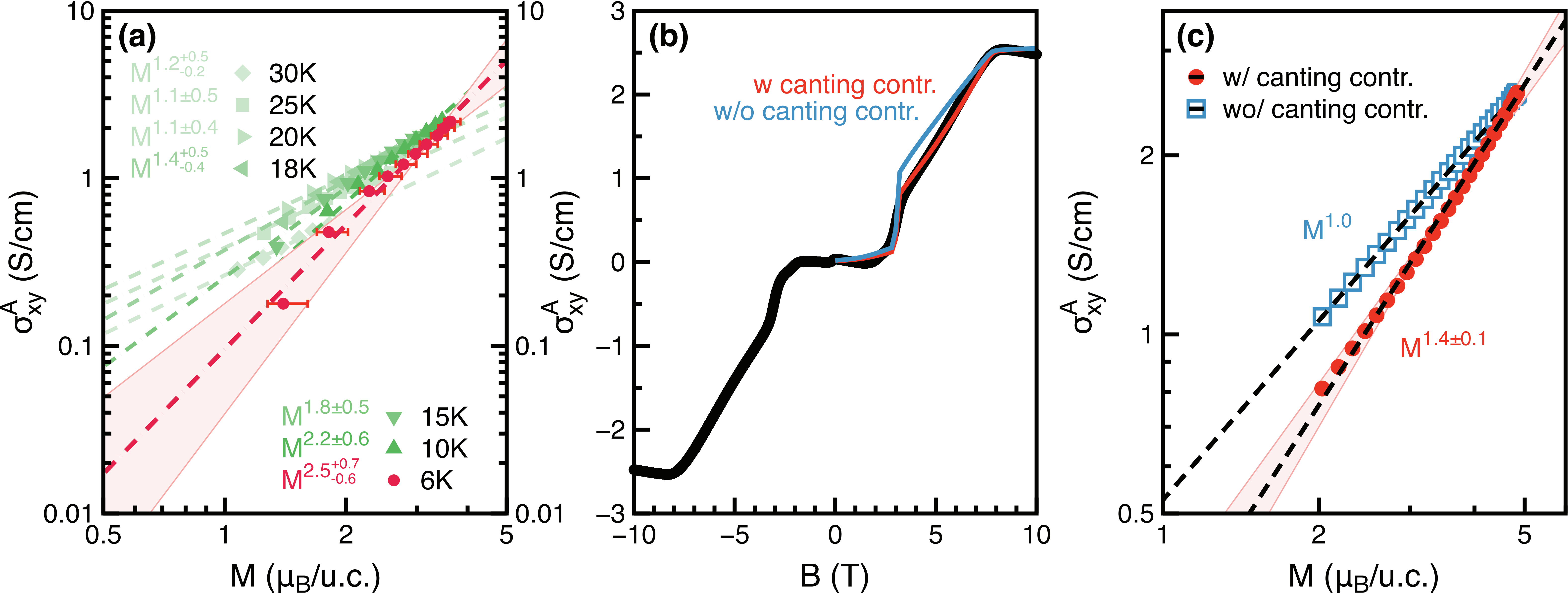

(a)
M^1.2 +0.5 -0.2
M^1.1±0.5
M^1.1±0.4
M^1.4 +0.5 -0.4
30K
25K
20K
18K
M^1.8±0.5
M^2.2±0.6
M^2.5 +0.7 -0.6
15K
10K
6K
σ^A_xy (S/cm)
M (μB/u.c.)
(b)
w canting contr.
w/o canting contr.
σ^A_xy (S/cm)
B (T)
(c)
w/ canting contr.
wo/ canting contr.
M^1.0
M^1.4±0.1
σ^A_xy (S/cm)
M (μB/u.c.)